\documentclass[10pt,a4paper,twocolumn,prl]{revtex4}
\usepackage{amsmath}
\usepackage{amsfonts}
\usepackage{amssymb}
\usepackage{graphicx}
\usepackage{epsfig}
\usepackage{dcolumn}
\usepackage{bm}

\begin{document}
\title{A Symmetry Breaking Model for X Chromosome Inactivation}
\author{Mario Nicodemi$^a$ and 
Antonella Prisco$^b$\\
$^a$ Dip. di Scienze Fisiche, Univ. di Napoli `Federico II', 
INFN, Via Cintia, 80126 Napoli,
Italy\\
$^b$ CNR Inst. Genet. and Biophys. `B. Traverso', Via P. Castellino 111, 80131 Napoli, Italy
}


\begin{abstract}
In mammals, dosage compensation of X linked genes in female cells 
is achieved by inactivation
of one of their two X chromosomes which is randomly chosen. 
The earliest steps in X-inactivation (XCI), namely the mechanism whereby
cells count their X
chromosomes and choose between two equivalent X, remain mysterious.
Starting from the recent discovery of X chromosome colocalization 
at the onset of X-inactivation, we propose a Statistical Mechanics model 
of XCI, which is investigated
by computer simulations and checked against experimental data. Our
model describes how a `blocking factor' complex is self-assembled
and why only one is formed out of many diffusible molecules,
resulting in a spontaneous symmetry breaking (SB) in the binding to
two identical chromosomes. 
These results 
are used to derive a scenario of biological
implications describing all current experimental evidences, e.g., 
the importance of colocalization. 
\end{abstract}

\maketitle

X chromosome inactivation (XCI) is the phenomenon in female mammal early embryo cells 
by which one of their two X chromosomes, randomly chosen, is 
transcriptionally silenced, and epigenetically inherited in
descendants, to equalize the dosage of X genes products
with respect to males \cite{Avner,Lucchesi,Brown}. 
Crucial aspects of this chromosome-wide stochastic regulatory
mechanism, necessary to survival, still elude comprehension despite
being the focus of substantial interest for their important scientific
and medical implication (see
\cite{Avner,Lucchesi,Brown} 
and Ref.s therein). 
Starting from the important discovery of X colocalization 
during XCI establishment \cite{Lee06,ap14}, 
in this paper we propose a Statistical Mechanics model of the early
steps of XCI. 

Actually, XCI is a multistep process involving \cite{Avner,Lucchesi,Brown}:
``counting'' the number of the X chromosomes of the cell, ``choice'' of the
inactive X, its silencing and maintenance. Silencing and maintenance 
start being understood: the former is induced by the action of the
{\em Xist} gene transcript, and maintenance of the inactive state
is a paradigm of epigenetic inheritance \cite{Brown,5}. 
Counting and choice are, instead, in many respects still mysterious, 
though it is well established they are controlled by yet unknown
sites located within a 1 Mb region on the X, the
X-chromosome-inactivation center ({\em Xic}),
containing several genes and regulators 
\cite{Avner,Lucchesi}, such as the {\em Xist} gene. 
We also know that cells having a normal number of autosomes (non sex
chromosomes) and extra copies of the X chromosome 
have only one active X, irrespective of the number of X's 
\cite{Avner,Lucchesi,Gartler}.  

This biological scenario suggests \cite{Avner,Lucchesi,Brown} 
that ``controlling factors" for counting and choice derive from autosomes
and interact with cis-acting regulatory sequences on the X chromosomes, 
whose position within the {\em Xic} is still unknown. Current models
postulate the existence of a ``blocking factor'' (BF) 
\cite{Avner,Lucchesi,Brown}, 
a complex made of X and autosomal factors, binding 
to the {\em Xic} of just one chromosome per
diploid cell preventing its inactivation, as the second unprotected
{\em Xic} in a female cell is inactivated by default. 
Multiple factors models were proposed as well \cite{Lee99,Lee05}. 
These models do not account, though, for the discovery
that colocalization, i.e., a physical proximity, occurs between
X chromosomes at the onset of XCI, specifically in
the {\em Xic} region, a phenomenon shown to be necessary for XCI 
\cite{Lee06,ap14}.
To comprehend the role of {\em Xic} colocalization, a description of the 
system is demanded that considers the influence of the spatial
configuration on the interaction of the {\em Xic} with the BF. 

The molecular nature of the blocking factor is itself still unknown: 
it must be a unique entity to perform
its function, though, several considerations (e.g., degradation,
over-production problems) exclude the possibility
that it is a single protein or RNA molecule. 
On the other hand, if a diffusible controlling factor is produced in
several copies that can statistically reach the target, asymmetric
binding to two equivalent chromosomes must be explained. 
While the BF can be envisaged as a unique complex formed
by autosomally derived molecules, why only one is formed is not understood.

We introduce a Statistical Mechanics schematic model of the diffusible
``controlling factors'' theory of X inactivation and we explain how a
supermolecular
complex can be self-assembled and why only one is formed out of
many molecules, resulting in a spontaneous symmetry breaking (SB) in
the binding to two identical chromosomal targets. 
We use, then, our new insights on 
the ``blocking factor" to derive biologically relevant implications
and depict a comprehensive scenario of experimental evidences 
that highlights the implications of {\em Xic} colocalization with respect
to the kinetics of XCI.

\begin{figure}[ht!!!]
\centerline{\hspace{0cm}\epsfig{figure=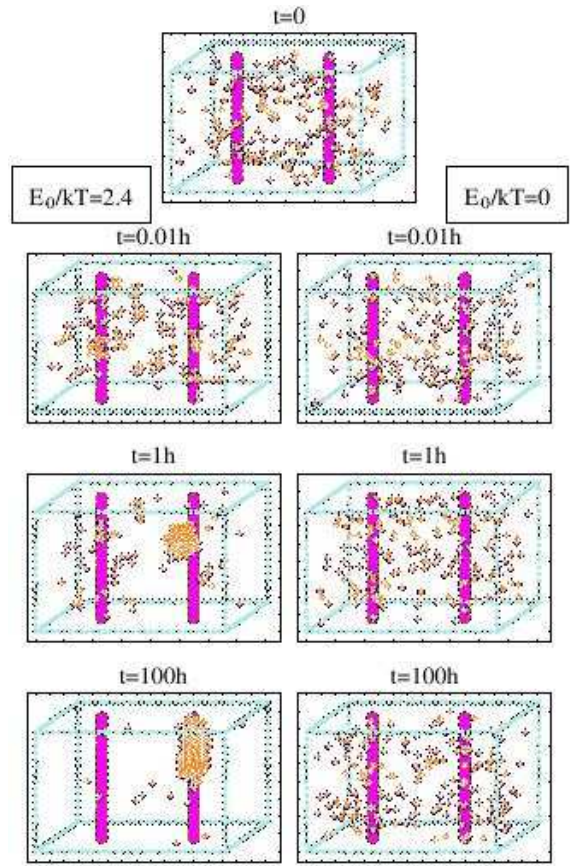,width=7.5cm,angle=0}}
\caption{\label{fig-plot3d} 
We show the evolution of our particle system, around two equally
binding ``chromosomes'', when the effective particle interaction energy
is $E_0=2.4kT$ ({\bf left}) and $E_0=0$ ({\bf right}), starting from the 
same initial random configuration at $t=0$. 
When $E_0=0$, a ``random walk''
diffusion is found. The evolution is drastically different for
$E_0=2.4kT$, where droplets of particles are formed, ending up in a
single cluster covering just one of the two equivalent chromosomes and,
thus, breaking their binding symmetry. 
}
\end{figure}

{\em Our model - }
In our model, for simplicity, we include just the essential components of the
process we are interested in (see Fig.\ref{fig-plot3d}): the two
relevant proximal portions of, say, the {\em Xic}, where
the diffusible controlling factors are assumed to bind, 
and a portion of space surrounding them. 
Such a volume includes an initially random distribution of molecular
controlling factors originated by autosomes. 
These factors are represented by diffusing particles 
having an affinity for their target regions on the two X chromosomes,
as well as a reciprocal affinity among themselves, as they can form a complex. 

In our schematic description, 
the two {\em Xic} segments are parallel, at a given distance $L$ 
in some units $d_0$ (of the order of the unknown molecular factors
size), in a volume (a cubic lattice with spacing $d_0$) 
of linear sizes $L_x=2L$, $L_y=L$ and $L_z=L$ around them 
(see Fig.\ref{fig-plot3d}). 
The diffusing factors randomly move from one to a nearest neighbor
vertex on the lattice. 
On each vertex no more than one particle can be present at a given
time \cite{nota_Ising}. 

Each particle interacts with its nearest neighbors via an effective 
energy $E_0$. Below, we mainly discuss the case where $E_0$ 
is of the order of a `weak' hydrogen bond energy, say 6 kJ/mole, which 
at room temperature corresponds to $E_0=2.4kT$ \cite{Watson} 
(the ``random walk'' model is recovered if $E_0=0$). 
The probability of a particle to move to a
neighboring empty site is proportional to the Arrhenius factor 
$r_0\exp(-\Delta E/kT)$, where $\Delta E$ is the energy change in the
move, 
$k$ the Boltzmann constant and $T$ the temperature  
\cite{Watson,Stanley}. The prefactor $r_0$ is the reaction
kinetic rate (setting the time scale here), 
depending on the nature of the molecular factors and of
the surrounding viscous fluid (for example, $r_0=30 sec^{-1}$ 
is a typical value of biochemical kinetics). 
Finally, since the {\em Xic} chromosome segments have an affinity for 
particles, each lattice site belonging to the chromosomes has a
binding energy $E_X$ (equal for the two X's) with particles; 
for simplicity we take $E_X=2.4kT$ too. 

The idea we illustrate below is that the molecular factors interaction,
$E_0$, induces cluster formation (see Fig.\ref{fig-plot3d}): when a
freely diffusing particle collides with a cluster of other particles,
it tends to ``stick'' to them, which produce cluster growth. We show
that if $E_0$
is above a given threshold, $E^*$ (of the order of `weak' hydrogen bonds), 
a phase transition occurs and the 
many clusters eventually coalesce in a single major ``complex''.
Interestingly, the time, $\tau$, to form the complex rapidly grows
with the X segments distance, $L$, explaining the important role of X
colocalization. 

{\em Computer Simulations - } 
We studied by Monte Carlo simulations \cite{Binder} 
the dynamics and the final state attained by the system. 
The size of our lattice is $L=16d_0$ (we checked our results for 
$L$ as large as $128d_0$), with periodic boundary conditions. 
Averages are over at least 256 runs 
and time is given in units of Monte Carlo lattice sweeps \cite{Binder}. 
The fraction of particles per lattice site 
in the examples
below (see Fig.\ref{fig-plot3d}) is $c=25 \cdot 10^{-3}$. 

Pictures of the 
system state at given time slices during a typical evolution are shown
in Fig.\ref{fig-plot3d}, which compares the simple ``random walk''
model ($E_0=0$) with the present model ($E_0/kT=2.4$). The
difference between the two is apparent: in the ``random
walk'' case, particles diffuse without forming any structure 
except for some binding to the `chromosomes'. 
When $E_0=2.4kT$, clusters of particles form, 
which end up in a single
big cluster covering only one of the `chromosomes'. 
This phenomenon, similar to nucleation where DNA acts as a seed
\cite{Frenkel}, illustrates how 
the formation of a single ``complex'' and the spontaneous
breaking of the binding symmetry between the two X chromosomes can occur. 
Notice that if the chromosome affinity tends to zero, $E_X\rightarrow 0$, 
a single cluster of particles is still eventually formed, but its binding to
the X's is unlikely.

\begin{figure}[ht]
\vspace{-1cm}
\centerline{\hspace{-3.25cm}\epsfig{figure=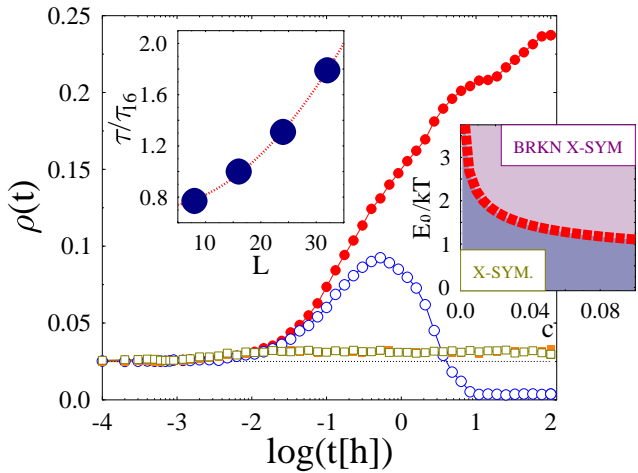,width=8cm,angle=-90}}
\vspace{-1.25cm}
\caption{\label{fig-dens} 
The average concentration around the chromosome on the left, $\rho_l$, 
and on the right, $\rho_r$, of Fig.\ref{fig-plot3d} are shown as a
function of time in a logarithmic scale. 
Circles refer to the case $E_0=2.4kT$ and squares to  $E_0=0$ 
(filled symbols for $\rho_l$, empty symbols for $\rho_r$). 
If $E_0=0$, the symmetry between the two chromosomes is
preserved during the evolution and $\rho_l(t)=\rho_r(t)$. 
When $E_0/kT=2.4$, the symmetry is broken: 
after a transient, $\rho_l$ grows one order of magnitude larger than
at the starting point and $\rho_r\rightarrow 0$. 
{\bf Right inset} The system phase diagram, derived by simulations, is shown in
the interaction energy-concentration plane, $(E_0,c)$, where the
broken curve fits the phase transition line. 
{\bf Left inset} We plot the time, $\tau$, to assemble the BF as a
function of the distance $L$ between the two X segments in a box of
given size $L_x=l_0$, $L_y=l_0/2$ and $L_z=l_0/2$ (with $l_0=64d_0$, 
$E_X=E_0=2.4kT$ and $c=25\cdot 10^{-3}$). 
Specifically, we show the ratio $\tau(L)/\tau_{16}$, where $\tau_{16}$
is the reference value for $L=16d_0$, the distance considered in the
rest of the paper. The 
fit is: $\tau=a+b L^2$. 
}
\end{figure}

Fig.\ref{fig-dens} shows the evolution, during the same kind of run 
of Fig.\ref{fig-plot3d}, of the average concentration around the
chromosome on the left, $\rho_l$, and on the right, $\rho_r$  
($\rho_l=N_l/V_l$ where $N_l$ is the number of particles in a cylinder, 
of radius $R=2.5d_0$ and volume $V_l=\pi R^2L_z$, centered around 
the left chromosome; analogously $\rho_r=N_r/V_r$). 
In the ``random walk'' case, $E_0=0$, the naive expectation from the
symmetry between the two chromosomes that $\rho_l(t)=\rho_r(t)$ 
is indeed verified at all times, $t$; 
$\rho_l(t)$ and $\rho_r(t)$ slightly grow up to a value comparable to the
initial random one. 
In the $E_0/kT=2.4$ case, $\rho_l$ and $\rho_r$ start from the
same initial value, $c$, but at some point one of the two has
a crisis as particles are all taken in the region of the other X whose
local concentration gets one order of magnitude larger than at the beginning.

The assembling of a single factor is attained when a balance 
is achieved between the entropy reduction and energy gain in the
process: for a given concentration of the particles, $c$, only when 
the interaction energy, $E_0$, is above the phase transition line 
value $E^*(c)$ (broken line in the phase diagram of the right inset in 
Fig.\ref{fig-dens}), 
a single major complex is formed and, thus, the original binding symmetry
between the two chromosomes is broken. 

{\em Discussion - } 
The SB mechanism we illustrated can explain why only one X chromosome 
per cell is active: it is the only one to recruit a sufficient number
of controlling factors. 
The time, $\tau$, to complete the assembling of the blocking factor
complex (BF) can be defined from the dynamics of the system ``order
parameter'', $m(t)=|\rho_l(t)-\rho_r(t)|/(\rho_l+\rho_r)$, which 
at long times is approximately exponential: $m(t)\propto \exp(-t/\tau)$. 
Alike Brownian processes, $\tau$ increases as the square of 
the X segment distance, $L$: $\tau\sim L^2$ (see left inset of 
Fig.\ref{fig-dens}). 
This suggests that only when the X's colocalize the BF can be
assembled in a time short enough to be useful on biological time scales.

Differences in the affinities of the X, $E_X$, induced for instance by
deletions of binding sites on one of them, result in a decisive
breaking of the symmetry in particle binding: the X with less 
affinity remaining ``naked'' and, thus, unprotected from inactivation. 
This explains biased XCI in embryonic tissues resulting 
from allelic differences in {\em Xic} sequences, 
(e.g., in the {\em Xce} locus \cite{21} 
or other regions \cite{Avner,Lucchesi,Brown}). 
XCI in female cells and lack of XCI 
in male cells could derive from a similar mechanism. 
With respect to XCI in polyploid cells, in our simulations the higher
is $c$ the larger the probability to have, at intermediate stages, two 
clusters, bound to two X, large enough to act as BF's. 
This could describe the stochastic 
nature of the `X chromosome/autosome ratio effect' 
\cite{Avner,Lucchesi,Brown}.

The SB model also rationalizes recent important deletion experiments across 
the {\em Xic} region, known to affect choice and counting 
(see \cite{ap1,Lee05,Clerc98,Lee99,la3,ap6,ap4} and Ref.s therein). 
We summarize here, in particular, the phenotype of three deletions,
which were instrumental in defining the role of the region 3' to 
{\em Xist} in counting and choice, namely 
$\Delta$65kb\cite{Clerc98}, 
{\em Tsix}$^{\Delta CpG}$ \cite{Lee99} and {\em Xite}$^{\Delta L}$\cite{ap6}.

The $\Delta$65kb deletion removes 65kb of DNA in the {\em Xic} region
relevant to the chromosome activation \cite{Clerc98}.
$\Delta$65kb causes non-random inactivation of the deleted X in
heterozygous XX cells\cite{Clerc98}, and X inactivation in XY cells\cite{ap4}. 
The explanation from BF models is that the
$\Delta$65kb deletion removes the binding sites for the blocking factor 
and the complex cannot bind the X any more. Interestingly, the X chromosome
bearing the deletion is not active, not even in male cells. 

The behavior of male cells is, however, drastically different in the case of
shorter deletions. 
The analysis of two smaller non-overlapping deletions within the 
above mentioned $\Delta$65kb sequence, namely the {\em Tsix}$^{\Delta CpG}$ 
deletion, removing the {\em Tsix} promoter,
and the {\em Xite}$^{\Delta L}$ mutation, removing {\em Xite}, added further
important information. 
In heterozygous XX cells, the {\em Tsix}$^{\Delta CpG}$deletion causes
non-random inactivation of the deleted X,
whereas in XY cells the {\em Tsix}$^{\Delta CpG}$ deleted X remains
active\cite{Lee99}. The {\em Xite}$^{\Delta L}$ phenotype is analogous 
\cite{ap6}. 
Finally, in homozygous {\em Tsix}$^{\Delta CpG}$ XX
mutants the choice of the active X is still random\cite{ap8}, but, 
importantly, in a fraction of cells both X's are inactivated 
(``chaotic counting'' \cite{Lee05}). 

\begin{figure}[ht!!]
\centerline{\hspace{-.25cm}\epsfig{figure=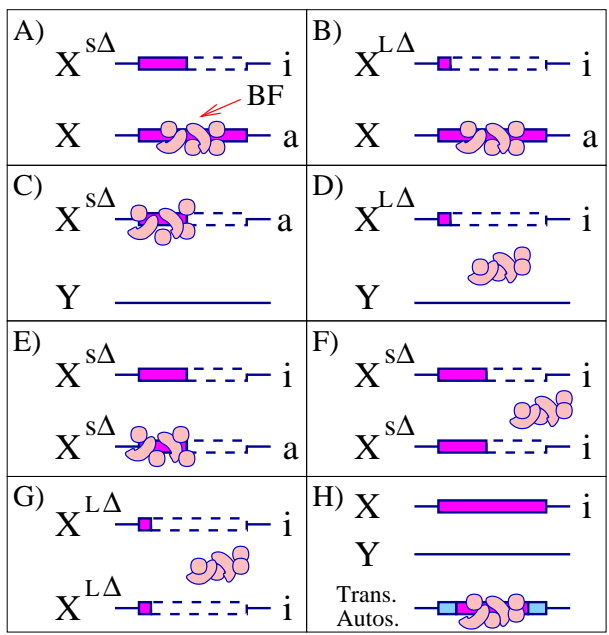,width=8cm,angle=0}}\vspace{0cm}
\caption{\label{fig-scheme_all} 
A pictorial summary of SB model results. 
Panels {\bf A,B)} consider heterozygous XX cells with either a comparatively 
long deletion in the {\em Xic} region (X$^{L\Delta}$), 
say $\Delta$65kb\cite{Clerc98}, or a shorter deletion (X$^{s\Delta}$), 
say {\em Tsix}$^{\Delta CpG}$ \cite{Lee99}. 
The mutated X, having a reduced overall affinity for the blocking
factor (BF), looses on average the competition for it ({\em skewed 
inactivation}). 
{\bf C,D)} In XY cells, the effects of X$^{s\Delta}$ and 
X$^{L\Delta}$ can be quite different: X$^{L\Delta}$ being unable to bind 
the BF. 
{\bf E,F)} In homozygous XX cells X$^{s\Delta}$, with a 
reduced affinity, succeeds in binding the BF only in a fraction of
cases ({\em ``chaotic counting''}\cite{Lee05}), 
whereas {\bf G)} X$^{L\Delta}$ is unable to bind BF. 
{\bf H)} In XY cells, transgenic autosomes with long enough {\em Xic}
insertions can bind the BF and the X is inactivated in a fraction of cases.
}
\end{figure}

As usual single BF models cannot explain these results 
\cite{Lee05}, the simulations with the SB version we propose
(summarized in Fig.\ref{fig-scheme_all}) allow a fresh look at the 
{\em Tsix}$^{\Delta CpG}$ and {\em Xite}$^{\Delta L}$ data. 
In our model, the blocking factor is a cluster of transacting factors 
which can bind several sites on a chromosome at the same time.  
The {\em Tsix}$^{\Delta CpG}$ deletion reduces the total blocking
factor/chromosome affinity. In our model, the difference in the affinity,
$E_X$, of the wild type and of the deleted 
chromosome explains, as described before, why choice is skewed in the 
heterozygous XX cells (see Fig.\ref{fig-scheme_all}A,B). 
At variance with the results from the longer 
$\Delta$65kb deletion, however, in the case of the smaller 
{\em Tsix}$^{\Delta CpG}$ and {\em Xite}$^{\Delta L}$ deletions, 
the mutated X remains active in XY cells. 
This is easily understood within the SB model: if 
binding sites are found in both the regions deleted
by {\em Tsix}$^{\Delta CpG}$ and {\em Xite}$^{\Delta L}$, 
then each mutation will reduce the affinity of the chromosome for the 
blocking factor, though neither deletion will fully erase the overall 
affinity. Thus, in XY cells the blocking factor can still bind the
deleted X chromosome (see Fig.\ref{fig-scheme_all}C,D), 
since there is no competing wild type X. 
Random choice in homozygous XX is explained as before 
(see Fig.\ref{fig-scheme_all}E). 
``Chaotic counting'' \cite{Lee05} in homozygous deleted
mutants derives from an analogous mechanism: deletions significantly
reduces the total X-chromosome/particles affinity 
(see Fig.\ref{fig-scheme_all}F,G) 
and, in a fraction of cells, the blocking factor doesn't bind at all. 

Transgene insertions into autosomes have also been analyzed 
\cite{21,Lee96,Ashworth}. 
When long {\em Xic} transgenes are introduced, in multiple copies
\cite{21}, into 
autosomes of male ES cells, inactivation of the single X occurs 
in a fraction of the cells \cite{Lee96,Ashworth}. 
In our view the mutated autosomes can 
bind the BF and compete with the X for it, leaving the real X
chromosome prone to inactivation (see Fig.\ref{fig-scheme_all}H).

The simple version of the SB model here discussed 
considers only a single kind of BF complex, though the model could easily
accommodate more than one (in a Potts-like variant), as recently 
proposed in \cite{Lee05}. 
It does not either imply that only one kind of soluble factors
is involved. 

Summarizing, the SB regulatory mechanism we propose 
describes the self-assembling of the blocking factor from diffusible 
DNA binding molecules and explains 
why only one is formed, i.e., the binding symmetry of
the two equivalent X's is broken. 
The emerging picture of its properties helps in delineating a 
scenario of biological implications reconciling within a single
framework the existing experimental evidences (e.g., X colocalization). 
In our model ``counting'' and ``choice'' are no longer 
distinct phenomena: they are regulated by the SB stochastic mechanism 
where time is an important parameter. 
More generally, the simplicity and robustness of the SB mechanism, 
illustrated here for XCI, suggest it can be widely 
used in random monoallelic expression processes \cite{Singh}.


\begin{thebibliography}{40}
\bibitem{nota_Ising} This version is a generalization of 
the Ising model \cite{Stanley}.
\end{thebibliography}
\end{document}